\documentclass[12pt]{article}
\usepackage{geometry}                		
\usepackage{graphicx}				
\usepackage{amsmath}								
\usepackage{amssymb}
\usepackage{amscd}
\usepackage{yfonts}
\usepackage{amsmath,amsthm,amssymb}
\usepackage{amsthm}
\newtheorem{theorem}{Theorem}

\newtheorem{definition}[theorem]{Definition}

 \newtheorem{lemma}[theorem]{Lemma}

\newcommand{\bv}{\bf v}
\newcommand {\bp} {\bf p}
\newcommand {{\bx}} {\bf x}
\newcommand {{\bk}} {\bf k}
\newcommand {\bP} {\bf  P}
\newcommand {\ba} {\bf a}

\date{}

\medskip

\title{Geometric and algebraic  approaches to quantum theory}
\author{Albert Schwarz}
\date{}							

\begin{document}
\author {  A. Schwarz\\ Department of Mathematics\\ 
University of 
California \\ Davis, CA 95616, USA,\\ schwarz @math.ucdavis.edu}


\maketitle


 \begin{abstract}
  {We show how to formulate physical theory taking as a starting point the set of states (geometric approach). We discuss the relation of this formulation to the conventional approach to classical and quantum mechanics and the theory of complex systems.  The equations of motion  and  the formulas for probabilities of physical quantities  are analyzed. A  heuristic proof of decoherence in our setting  is  used to justify the formulas for probabilities.  We show that  any physical theory theory can be obtained from classical theory if we restrict the set of observables. This remark can be used to construct models with any prescribed group of symmetries; one can hope that this construction  leads to new interesting models that cannot be build in the conventional framework.
  
  The geometric approach can be used to formulate quantum theory in terms of Jordan algebras, generalizing the algebraic approach to quantum theory. The scattering theory can be formulated in geometric  approach. }
  \end {abstract}
  
  \section {Introduction}
Let us start with some  very general considerations.

Almost all physical theories are based on the notion of state  at the moment $t.$ 
The set  of states will be denoted  ${\cal C}_0$.{\it  We assume that  the set of states is a convex set.} This means
we can consider   mixtures of states:  taking states $\omega_i$ with probabilities $p_i$ we obtain the mixed state denoted $\sum 
p_i\omega_i$. Similarly if we have a family of states $\omega({\lambda})$ labeled by elements of a set $\Lambda$ and a probability distribution  on $\Lambda$ ( a  positive measure $\mu$ on $\Lambda$ obeying $\mu(\Lambda)=1$) we can talk about the mixed state $\int_{\Lambda} \omega (\lambda) d\mu.$ 

{\it  We assume that  the set of states  ${\cal C}_0$ is a subset of a topological linear space $\mathcal {L}.$ } The extreme points  of ${\cal C}_0$ are called pure states.  (In other words a state is pure if it cannot be represented as a mixture of two distinct states.)  We assume that for all convex sets we consider every point is a mixture of extreme points.
( If a set is a convex compact subset of locally convex topological vector space this assumption is a statement of Choquet-Bishop-de Leeuw theorem.)

Our assumptions are valid both  for classical and quantum mechanics.

Instead of the set $\mathcal {C}_0$ we can work with the corresponding cone  \footnote {We define a cone (more precisely a convex cone) as a set  that for every point $x$ contains  the point $Cx$ where $C$ is a positive number. The dual cone  is defined as a set of linear functionals that are non-negative on the cone. }
 $\cal C$  (the set of points of the form $\alpha x$ where $\alpha$ is a positive real number,  $x\in \mathcal {C}_0$). The elements of this cone are called non-normalized states. Two non-zero elements $x,y\in \cal C$ determine the same normalized state if they are proportional :
$y=Cx$ where $C$ is a positive number. ( To identify  the cone $\cal C$  factorized with respect to this equivalence relation with ${\cal C}_0$ we should assume that ${\cal C}_0$ does not contain zero and every ray $\{Cx\}$ where $C>0$ contains at most one point of ${\cal C}_0.$)

An observable  specifies a linear functional  $a$ on ${\cal L}.$ This requirement agrees with the definition of mixed state: if $\omega=\sum p_i\omega_i$ then $a(\omega)$ is an expectation value of $a(\omega_i)$. (The non-negative numbers $p_i$ obeying $\sum p_i=1$ are considered as probabilities.)

We consider deterministic theories. This means that  the state in the moment $t=0$ (or in any other moment $t_0$)  determines the state in arbitrary moment $t$. Let us denote by $\sigma (t)$ the operator  transforming the state in the moment $t=0$ into the state in the moment $t$ (the evolution operator). The evolution operators constitute a one-parameter family $\sigma(t)$ of invertible maps $\sigma (t): \mathcal {C}_0\to \mathcal {C}_0$ , that can be extended to linear maps of $\cal L.$  In other words the operators $\sigma(t)$  belong to the group $\cal U$ of automorphisms of $\mathcal {C}_0$ (to the group  of linear bicontinuous maps of $\cal {L}$ inducing invertible maps of $\mathcal {C}_0$ onto itself). In some cases one should impose an additional condition
$\sigma(t)\in \mathcal {V}$ where $\cal V$ is a subgroup of $\cal U.$

{\it We describe a physical  theory fixing a bounded convex set ${\cal C}_0$ and a subgroup $\cal V$ of the group of automorphisms  of this set. Imposing some technical conditions that are valid in the case of quantum mechanics and wrong  for classical mechanics we prove a generalization of decoherence and derive  formulas for probabilities generalizing the formulas of quantum mechanics.}

The evolution operator satisfies the equation
\begin {equation}
\label {EM}
\frac {d\sigma}{dt}=H(t)\sigma (t)
\end {equation}
(equation of motion).
Here $H(t)\in Lie(\cal V)$  (the  "Hamiltonian") is a $t$-dependent  element of the tangent space to the group $\cal V$  at the unit element.  ( We will use the name " infinitesimal automorphism" for an  element of $Lie (\cal V).$) \footnote { Knowing the topology in $\cal L$ we can define in various ways the topology in $\cal V\subset \cal U.$
This allows us to define $Lie (\cal V)$ as the tangent space at the unit element.  Alternatively we can define
$Lie (\cal V)$ as a set of all linear operators $h$  such that the equation $\frac {d\sigma}{dt}=h\sigma (t)$
has a solution obeying $\sigma(t)\in {\cal V}, \sigma(0)=1.$  Notice that in the case when $\cal V$ is
infinite-dimensional $Lie (\cal V)$ is not necessarily a Lie algebra. We disregard the subtleties appearing in infinite-dimensional case.}

The equation (\ref{EM}) can be regarded as a definition of $H(t).$ However, usually we go in opposite direction: the physical system we  consider is specified by the operator $H(t)$ (by the equation of motion) and our goal is to calculate the evolution operator solving the equation of motion.

Examples

{\it Classical mechanics}

The cone $\cal C$ consists of positive measures $\mu$ on symplectic manifold$M$, we assume that $\mu (M)<\infty$

 The set ${\cal C}_0$  consists of probability distributions (normalized positive measures, $\mu (M)=1$).

Observables are functions on symplectic manifold

$\cal V$ is  the group of symplectomorphisms

$Lie (\cal V)$ is the Lie algebra of Hamiltonian vector fields

The equation of motion is the Liouville equation $\frac {d\rho}{dt}=\{H,\rho\}$ where $\rho$ stands for the density of the measure and $\{\cdot,\cdot\}$ denotes the Poisson bracket.

{\it Quantum mechanics}

The  set ${\cal C}_0$ consists of density matrices ( positive trace class operators in  real or complex Hilbert  space $\cal H$ having  unit trace: $Tr K=1$).  Omitting the condition $Tr K=1$ we obtain the cone $\cal C$.\footnote {Physicists always work in complex Hilbert space imposing the condition of reality on the states if necessary. We prefer to work in real Hilbert space.} Notice that $\cal C$ is a homogeneous self-dual cone; such cones are closely related to Jordan algebras ( see Section 4).

In textbook quantum mechanics it is assumed that that $\cal H$ is a complex Hilbert space. In this case observables are identified with self-adjoint operators.
A self-adjoint operator  $\hat A$   specifies a linear functional by the formula
$K\to Tr \hat AK.$

$\cal V =\cal U$ is isomorphic to the group of invertible isometries. (They are called orthogonal operators if the Hilbert space is real and unitary operators if the Hilbert space is complex.) An operator $\hat V\in \cal U$  acts  on $\cal C$ and on $\mathcal {C}_0$  by the formula $K\to \hat VK \hat V^*$. 

The tangent space of  this group at the unit element  can be regarded as the Lie algebra $Lie (\cal U)$; it  consists of bounded operators obeying $\hat A+\hat A^*=0$ (skew-adjoint  operators).
( In complex Hilbert space  instead of skew-adjoint operator $\hat A$ we can work with self-adjoint operator $\hat H=i\hat A$.)
\footnote {The group $\cal U$ can be considered as Banach Lie group; the tangent space to it can be defined as the set of tangent vectors to curves (to  one-parameter families of operators) that are differentiable with respect to the norm topology. Notice, however,  that the families of evolution operators appearing in physics usually do not satisfy this condition (but they can be approximated by differentiable families).
 For a self-adjoint operator $\hat H$ in complex Hilbert space the operator $\hat A=-i\hat H$ can be considered as a tangent vector to the one-parameter group of unitary operators $e^{{\hat A}t}=e^{-i{\hat H}t}.$ This family is differentiable in norm topology only if the operator $\hat H$ is bounded. We can define $ Lie (\cal U)$ as a set of all self-adjoint operators, but in this definition $Lie (\cal U)$ is not a Lie algebra (a commutator of self-adjoint operators is not necessarily well defined).  Problems of this kind appear also in geometric approach to quantum theory; we neglect them.}

The equation of motion can be written in the form

$$\frac {dK}{dt}=A(t)K=-\hat A(t)K+K\hat A(t).$$
where $\hat A(t)$ is a family of skew-adjoint operators in $\cal H$.

In complex Hilbert space we can write the equation of motion as follows:

$$\frac {dK}{dt}= H(t)K=i(\hat H (t) K-K\hat H(t))$$
where $\hat H(t)$ is a family of self-adjoint operators ( here $\hat H(t)=i\hat A(t)$).

{\it Quantum theory in algebraic approach}

The starting point is a unital associative algebra  $\cal A$ with involution $^*$

The set of states ${\cal C}_0$  is defined as the space of positive normalized linear functionals on $\cal A$ (the functionals obeying $\omega (A^*A)\geq 0, \omega (1)=1$). 


$\cal V=\cal U$ denotes the group of involution preserving automorphisms of $\cal A$ (they act naturally on states)

To relate the algebraic approach to  the Hilbert space formulation we use the GNS (Gelfand-Naimark-Segal)  construction: for every state $\omega$ there exists
pre Hilbert space $\cal H$, representation $A\to \hat A$  of $\cal A$ in $\cal H$ and a cyclic vector $\theta \in \cal H$ such that
$$\omega (A)=\langle \hat A\theta,  \theta\rangle $$
( We say that vector $\theta$  is cyclic if every vector $x\in \cal H$ can be represented in the form $x=\hat A \theta$ where $A\in \cal A.$ Notice that instead of pre Hilbert space $\cal H$ one can work with its completion, Hilbert space $\overline {\cal H}$, then $\theta$ is cyclic in weaker sense: the vectors $\hat A\theta$ are dense in Hilbert space. )

{\it Complex systems}

Let us consider a family of  physical theories depending on parameter $\lambda\in \Lambda$ . This means that the set of states ${\cal C}_0$, the group $\cal V$ and the "Hamiltonian" $A$ depend on this parameter. A state of complex system  is  a   probability distribution on families  of states  $\omega (\lambda)\in {\cal C}_0(\lambda)$,  the evolution  of state is governed by random "Hamiltonian" $A(\lambda)$ (see, for example, \cite {PAR}).

A typical example of complex system is spin glass  \cite {SPI}. In this case
${\cal C}_0$ does not depend on parameter, its elements are states of classical or quantum system of spins.  The evolution is governed by random Hamiltonian; for example one can take a quantum Hamiltonian that is quadratic with respect to spins; the coefficients of the quadratic expression are random parameters.

Notice that the  random "Hamiltonians" in  the theory of  complex  systems are time-independent. In the proof of decoherence  we use time-dependent random "Hamiltonians".

\vskip .2in
 In Section 2 we describe the approach to physical theory where the primary notion is the convex set of states (geometric approach). We discuss its relations to algebraic approach and to textbook quantum mechanics. We show that a generalization of decoherence is correct in geometric approach and use this statement to derive the formulas for probabilities from the first principles.  To prove decoherence we study the interaction of physical system with random environment that is modeled as random adiabatic perturbation of the equations of motion.   In Section 3 we show that physical theory can be obtained from classical theory if we restrict the set of observables. ( In classical theory a state can be represented as a mixture of pure states in unique way. If not all observables are allowed some states  should be identified. This identification permits us to construct any theory from classical theory.) In Section 4 we  analyze the relation to Jordan algebras . \footnote{The set of self-adjoint elements (observables) is not closed 
under multiplication, but it is closed with respect to the operation 
 $a\circ b=\frac 1 2 (ab+ba)$. This remark led to the notion of Jordan algebra ( axiomatization of this operation).  Jordan algebras  can be regarded as the natural framework of algebraic approach. This statement is prompted  by the 
Alfsen-Shultz theorem: Cones of states of two $C^*$-algebras are isomorphic iff corresponding Jordan algebras are isomorphic.} In Section 5   that does  not depend on Sections 2,3,4 we discuss the notions of particle
and quasiparticle in algebraic and geometric approaches.  In this discussion it is useful to take as a starting point the cone $\cal C$ of non-normalized states.
The set of all endomorphisms of $\cal C$ (the set of all linear continuous operators in $\cal L$ mapping ${\cal C}$ into itself)  will be denoted by $End (\cal C).$  This set is a semigroup with respect  to  composition of operators; it is also closed with respect to addition and with respect to multiplication by a positive scalar (it is a semiring)\footnote { Usually semiring is defined as set with operations of addition and multiplication obeying obeying the standard axioms (associativity, distributivity, existence of $0.$).We include the multiplication by a positive scalar in the definition of semiring.}  One of the ways to  develop scattering theory is based on consideration of the subset $\cal W$ of $End (\cal C)$ that is also closed with respect to these operations and with respect to the action of elements of $\cal V$.  (Notice that the automorphisms of $\mathcal {C}_0$ act naturally on $\cal C$ specifying elements of  $End ({\cal C}).$)

In examples $Lie (\cal V)$ can be identified with the set of global observables, the semiring $\cal W$ consists of quasilocal endomorphisms (quasilocal observables).   

Working in the framework of the algebraic quantum theory we define the data of geometric approach taking as $\cal L$  the algebra $\cal A$ considered as a linear space or the space $\cal A^*$ of linear functionals on $\cal A.$ The cone in $\cal A$ is spanned by $A^*A$ where $A\in \cal A$,  the cone  $\cal C$ in $\cal A^*$ is dual to this cone (this is the cone of positive linear functionals on $\cal A$, i.e. the cone of linear functionals obeying $f(A^*A\geq 0$). Automorphisms of algebra specify automorphisms of cones; we take $\cal V$ as the group of these automorphisms.  The map $A\to B^*AB$ where $B\in \cal A$ specifies an endomorphism of the cone in $\cal A$, the dual map is  an endomorphism of the dual cone. We define $\cal W$ as a semiring generated by these maps.
Every element $B\in \cal A$ specifies two operators on the space ${\cal A}^*$: for a linear functional $\omega$ on $\cal A$ we define $(B\omega)(A)=\omega(AB), (\tilde B\omega)(A)=\omega(B^*A).$ The endomorphism of $\cal C$ described above can be written as $\tilde B B.$

To define (quasi)particles we need the action of space and time translations.
 We define particles as elementary excitations of ground state and quasiparticles as elementary excitations of any translation-invariant stationary state (see Section 5 for details). 
 
 We relegate the scattering theory to follow up papers \cite {GA2},\cite {GA3}, \cite {GA4}. These papers are at least formally independent  of the present paper. The paper \cite {GA2} is devoted to the scattering theory in algebraic approach; it generalizes the results of \cite {SC} (see also the Chapter 13 of  \cite {MO}. The main goal of \cite {GA2} is to provide a convenient way to compare the constructions of \cite {GA3} with standard constructions. To achieve this goal we consider algebras over real numbers in this paper.  In the paper \cite {GA3} we develop the scattering theory in geometric approach. Notice that in this approach the conventional scattering matrix cannot be defined , but there exists a very natural definition of inclusive scattering matrix (see \cite {S},\cite {SC}, \cite {MO} ). The paper \cite {GA4} is devoted to scattering theory in the framework of Jordan algebras; it relies on the definitions given in \cite {GA3}, but is mostly independent of \cite {GA2},\cite {GA3}.
 
 We do not perform any calculations in this paper. It is natural to ask whether the geometric approach is convenient for calculations. The answer to this question is positive. The inspiration for geometric approach came in part from the formalism of $L$-functionals \cite {SCH},\cite {T},
 \cite {S}, \cite {MO} ( see also Section 3).  In this formalism one works with the space of states over Weyl or Clifford algebra. One can construct diagram techniques of perturbation theory in this formalism; they can be used not only in quantum field theory, but also in equilibrium and non-equilibrium statistical physics. These diagram techniques  are equivalent to diagrams of Keldysh formalism and of thermo-field dynamics (see for example \cite{CU}, \cite {K}) that were applied to many problems of statistical physics. The same diagrams are useful also in calculation of inclusive scattering matrix (see \cite {S}, \cite {II} for more detail).  The formalism of $L$-functionals can be used also  in the theory of complex systems, in particular in the theory of spin glasses. It is important to notice that in this formalism (as well as in Keldysh formalism)  the replica trick is not needed.

\section {Geometric approach to  physical theories} 

{\it We start with a bounded convex closed  set ${\cal C}_0\subset \cal L$ and  a subgroup $\cal V$ of the automorphism group $\cal U$ of ${\cal C}_0.$ }

Here $\cal L$ is a normed space over $\mathbb {R}$  and automorphisms of ${\cal C}_0$ are by definition invertible linear bicontinuous operators in $\cal L$ mapping ${\cal C}_0$ onto itself. (More generally we can assume that $\cal L$ is a linear topological space.)

We assume that one can obtain the evolution operator $\sigma (t)\in \cal V$   acting in ${\cal C}_0$ from an  operator $A$ (we call it "Hamiltonian") using the equation of motion :
\begin {equation} \label {EMO} \frac {d\sigma}{dt}=A\sigma (t)
\end {equation}
(We say that $A$ is an infinitesimal automorphism if this equation has a solution $\sigma(t)\in \cal  V$. The set of all  infinitesimal automorphisms is denoted $Lie (\cal V).$)

We will consider also a more general case when $A$ in (\ref {EMO}) depends on $t$ (the "Hamiltonian" is time-dependent). 
 
  Let us  assume that there exists a basis of the complexification of $\cal L$ consisting of eigenvectors of $A$. We say in this case that $A$ is diagonalizable. Let us denote by $( \psi_j)$ the basis consisting of eigenvectors:
$$ A\psi_j=\epsilon_j \psi_j$$
 It follows from the boundedness of ${\cal C}_0$ that $\epsilon_j\in i\mathbb {R}.$ Notice that the boundedness implies also that $A$ does not have non-trivial Jordan  cells (there are no associated eigenvectors). This means that we should expect that the operator $A$ having only point spectrum is diagonalizable.
 
 Example: textbook quantum mechanics. As we noticed in this case $\mathcal {C}_0$  consists of density matrices and the equation of motion in the case of time-independent Hamiltonian has the form 
 $$\frac {dK}{dt}=AK=i(\hat H K-K\hat H).$$ The space $\cal L$ consists of all self-adjoint operators belonging to trace class and the complexification of this space 
consists of all operators belonging to trace class. The cone $\cal C$ consists of positive trace class operators.

Let us assume that $\hat H$ is diagonalizable. Then there exists an orthonormal basis $\phi_n$ of the Hilbert space  consisting of eigenvectors  of $\hat H$ with eigenvalues  $E_n$. The "Hamiltonian"  $A$  also is diagonalizable  wiith  eigenvectors denoted by $\phi_{mn}$;
eigenvalues of the "Hamiltonian" $A$ are differences of eigenvalues of $\hat H$ ( up to a factor of $i$.)

In $\hat H$-representation  the operator $\hat H$ is a represented by a diagonal matrix with diagonal entries $E_n$; the "Haniltonian" $A$ acts in the space of matrices   and the eigenvectors $\phi_{mn}$ are matrices having only one non-zero element equal to $1.$ Every density matrix $K$  can be represented   in the form $K=\sum k_{mn}\phi_{mn}$; we say that $k_{mn}$ are matrix entries of $K$ in $\hat {H}$- representation.

One can prove that the interaction with random environment leads to  vanishing of non-diagonal matrix entries of density matrix  in $\hat H$-representation. The diagonal entries do not change. This effect is known as decoherence;
it corresponds to the collapse of the wave function in old axioms of the theory of measurements. The decoherence is usually derived from the interaction of  of a quantum system with a large thermal base. However, it  can be shown that random adiabatic interactions  also lead to decoherence. We will sketch of proof of this fact  (see \cite{TSC}, \cite{MO} for details).  \footnote {It seems that in reality decoherence  in a quantum system often  comes from random  adiabatic interactions, mostly with  electromagnetic  fields  generated by  objects nearby (as in experiments confirming decoherence). This  fact is not important for us, however it supports  our choice of the model describing the interaction with environment.  Notice that non-adiabatic interactions also can kill non-diagonal entries of density matrix, however these interactions in general change diagonal entries (hence strictly speaking they do not lead to decoherence). Another  proof of decoherence is based  on semiclassical considerations.  This proof is similar from the mathematical viewpoint to our proof; it is also based on consideration of   fast oscillating integrals, but  the large phase comes in this case form small Planck constant (in our proof it comes from adiabatic parameter). It is important that our considerations are not related in any way to the classical limit  or to the theory of measurements.}

 To give a proof we include the Hamiltonian $\hat H$ in a smooth family of Hamiltonians $\hat H(g)$ assuming that $\hat H(0)=
\hat H(g_0)=\hat H.$ We assume that  there exist  eigenvectors $\phi_n (g)$ of Hamiltonians $\hat H(g)$ depending smoothly on $g$  and obeying $\phi_n(0)=\phi_(g_0)=\phi_n.$
The family of Hamiltonians $\hat H(at)$ where $a$ is a small parameter is an adiabatic (=slow varying) family.
It is easy to check that in adiabatic approximation
    $$e^{-i\frac {1}{a}C_n(g)}\phi_n(g)$$
where $g=at$ ,  $\frac {dC_n(g)}{dg}=E_n(g)$ ,obeys Schr$\ddot {o}$dinger equation for the time -dependent Hamiltonian $\hat H(at)$ (we neglect terms tending to zero when the adiabatic parameter $a$ tends to $0$). This allows us to  represent the evolution of the density matrix  in $\hat H(at)$-representation by  the formula
$$K(t)=\sum_{m,n} k_{mn}(t)\phi_{mn}(t)$$
where 
$$k_{mn}(t)= e^{-i \frac {1}{a}C_{mn}(at)},  \frac {dC_{mn}(g)}{dg}=E_m(g)-E_n(g).$$
We see that the diagonal  matrix entries do not change and the non-diagonal entries acquire a phase factor with large phase. 

Let us suppose now that  $H(g)$ is  a random Hamiltonian. Under certain conditions on the probability distribution
one can check that the expectation value on non-diagonal entries vanishes in the limit $a\to 0.$ This means that the interaction with environment described by random adiabatic Hamiltonian leads to decoherence.

Let  us show that very similar consideration allows us to prove decoherence in geometric approach. We consider  
the evolution  governed by the equation (\ref {EMO}) where $A$ is a diagonalizable operator with eigevalues $\epsilon_j\in i\mathbb{R}$ and eigenvectors $\psi_j.$
 
  Let us suppose that $A(g)$ is a smooth   family  of elements of  $Lie (\cal V)$ and $A(0)=A$.  Then  for a right choice of the basis $(\psi_j)$ and for  $|g|<\delta_j$ we  can construct  vectors $(\psi_j(g))$ that depend smoothly on $g$ in such a way that 
 \begin {equation}\label {DI} A(g) \psi_j(g)=\epsilon_j (g)\psi_j(g)
 \end {equation}
 where  $\psi_j(0)=\psi_j.$
 (We assume that  all non-zero eigenvalues of $A(g)$  are  at most finitely degenerate)

  Notice that in general  the basis $\psi _j(g)$ cannot be extended continuously to all $g$; we assume that there exists a piecewise continuous ( and piecewise smooth ) extension of this basis for all $g.$

  {\it We say $\psi_j$ is a robust zero mode of $A$ (= robust stationary state) if $\epsilon_j(g)=0.$}

 Let us model the interaction with environment by random adiabatic  (=slowly changing) " Hamiltonian"  $A(g(t))$  ("adiabatic" means that we can neglect the derivative $\dot g(t))$). \footnote { We want to use this model to prove decoherence in our setting. As we show it is sufficient to  consider only adiabatic interactions to explain decoherence}.  
 Then in the adiabatic approximation the evolution of the vector $\psi_j$ is described by the formula
 $$ \sigma (t)\psi_j= e^{\rho_j(t) }\psi_j(g(t)),$$
 where $\frac {d\rho_j}{dt}=\epsilon_j(g(t)).$ ( In other words  $\sigma(t)\psi_j$ obeys the equations of motion if we neglect $\dot g(t)$.) The phase factor is equal to $1$  for robust zero modes, the phase of this factor is large for all other modes.
 
Imposing some conditions on the random "Hamiltonian" $A(t)$ one can prove that {\it in average the random phase factors
 $e^{\rho_j(t)}$ vanish unless $\phi_j$ is a robust zero mode of } $A$.

  Let us introduce the projection $P'$ of $\cal L$ onto the subspace spanned by robust zero modes of $A.$ ( We say that $P'\phi_j=0$ if $\phi_j$ is not a robust zero mode.). We can say that  under certain conditions  the random adiabatic  "Hamiltonian" $A(t)$ kills all modes except robust zero modes (= transforms $x\in {\cal C}_0$ into $P'x.$). In textbook quantum mechanics  this means that random adiabatic Hamiltonian  $\hat H$ kills  all  non-diagonal entries of density matrix $K$ in $\hat H$-representation and does not change diagonal entries.  This effect is known as decoherence. Hence we proved an analog of decoherence in our approach.
  
   We will prove that $P'x\in {\cal C}_0$ if $x\in {\cal C}_0.$ The 
   set $P'( {\cal C}_0)$ is the set of states that are robust zero modes(=robust stationary states); we say that extreme points of this set are pure robust zero modes (=pure robust stationary states).
  
   If $x\in {\cal C}_0$ we represent $P'x$ as a mixture of pure robust zero modes $z_k$ with  coefficients $p_k$ :  $$P'x=\sum p_k z_k.$$   As usual the numbers $p_k$
  should be interpreted as probabilities.  
  
  In general the coefficients $p_k$
are not defined uniquely. This situation is familiar in quantum mechanics:  density matrices can be represented  as  mixtures of pure states in different ways. However, in conventional quantum mechanics generically a representation of $P'x$  as a mixture of pure robust zero modes is unique.  (A generic Hamiltonian has only simple eigenvalues.  In this case every density matrix commuting with Hamiltonian has a unique representation as a mixture of stationary states corresponding to eigenvectors of the Hamiltonian.)

Notice that our considerations show that after averaging with respect to adiabatic random perturbations every state becomes a stationary state. This statement is consistent with quantum mechanical results ( a density matrix having only diagonal entries in $\hat H$-representation describes a stationary state).
However, this statement cannot be general.  One of the reasons why  our considerations were not universal was a very strong condition we imposed on robust zero modes: we   said the $\psi_j$ is a robust zero mode if  for {\it all} $g$ there exists a  zero mode $\psi_j(g)$ of the "Hamiltonian" $A(g)$ depending continuously on $g$ and coinciding with $\psi_j$ for  $g=0$.  It is more natural to 
impose this condition not for all $g$, but only for small $g.$
In what follows we use this modification of the definition of robust  zero mode (in a little bit  weaker form:  we require the existence of  zero mode of $A(g)$ that tends to $\psi_j$  as $g\to 0$).

Notice that to prove decoherence with the modified definition of robust zero mode  we should restrict ourselves to small adiabatic perturbations of the original "Hamiltonian".

One can apply the above considerations to the case when $A$ is an arbitrary infinitesimal automorphism (not necessarily the "Hamiltonian").  If $Ax=0$ ( i.e. $x$ is a zero mode of $A$) we say that $x$ is a robust zero mode if 
for every infinitesimal automorphism $A'$ in a small neighborhood of $A$ we can find a zero mode $x'$ in a small neighborhood of $x.$

 Let us define an observable as a pair $(A,a)$ where $A$ is an infinitesimal automorphism and $a$ is an $A$-invariant linear functional  on the space of states; it  has  a physical meaning of the expectation value of the observable.  ( We say that a linear functional $a$ is $A$-invariant if $a(Ax)=0$ for all $x\in \cal L.$)  For example for energy $(H, h)$ the infinitesimal automorphism $H$ is the "Hamiltonian" and $h$ is the expectation value of the energy. The state with minimal value of $h$ is a ground state.
 
Let us denote by $(Ker A)_r$ the space of robust zero modes of $A$ and by $P'$ the projection  $P':\mathcal {L}\to (Ker A)_r$  sending  all eigenvectors of $A$ that are not robust zero modes to zero.  Pure robust zero modes are defined as extreme points of the set $P'({\cal C}_0).$
To calculate probabilities of $A$ in the state $x$ we should represent  the robust zero mode $P'x$  where $x\in {\cal C}_0$ as a mixture of robust pure zero modes: $P'x=\sum p_k z_k.$ Then $p_k$ {\it  can be considered  the probability to find the the value $a(z_k)$  when we measure}$A$. (We assume that the numbers $a(z_k)$ are different. If this condition is not satisfied we should 
calculate the probability to obtain the value $\alpha$  summing all $p_k$ with $a(z_k)=\alpha.$)

 In the textbook quantum mechanics  we take $A$ as  a commutator with a self-adjoint operator $\hat A$ multiplied by $i$  and define $a(K)=Tr \hat A K$. Notice that $a(K)$ is not necessarily finite; for example, in translation-invariant state the value of energy is in general infinite (but we can talk about the density of energy and about the difference of energies). The situation in the geometric approach is similar.
 
We started with a proof of decoherence in quantum mechanics and noticed that the same proof works in much more general situation. Let us show how the decoherence in quantum mechanics can be derived from general results.
 
 We already noticed  that in $\hat A$-representation the basis of eigenvectors of $A$ consists of matrices having only one non-zero entry equal to $1.$   If a density matrix can be represented in the form $f(\hat A)$ for some function $f$ then the corresponding state is a robust zero mode of $A$ (if $\hat A'$ is close to $\hat A$  then $A'$ has a  zero mode $f(\hat A')$ that is close to $f(\hat A)$ ).  All operators of the form $f(\hat A)$ are diagonal in $\hat A$-representation. If  all eigenvalues  of $\hat A$ are simple then all   density matrices that are diagonal in $\hat A$-representation have the form $f(\hat A)$ hence they specify robust zero modes.

 The projection $P'$ in $\hat A$-representation sends every density matrix into  diagonal matrix (decoherence, collapse of wave function). 
 
 There are many choices for the functional $a$ for given infinitesimal automorphism $A.$ However,
 most of them are equally good. Let us consider , for example, the infinitesimal time translation $H$ and the corresponding "energy" functional $h$. Then decoherence gives a projection on the robust part of $Ker H.$ Let us consider the textbook quantum mechanics. Then the pure states in $Ker H$
 are precisely the eigenstates $\phi_n$ of the Hamiltonian $\hat H$ with eigenvalues $E_n$; we suppose that all $E_n$ are distinct. With the natural choice of $h$ we have $h(\phi_n)=E_n.$ With any other choice $h'$  of "energy" functional we have $h'(\phi_n)=f(E_n)$ for some function $f.$ This means that  knowing the probabilities for the natural choice we can calculate the probabilities for $h'.$ Conversely, if $f$ is increasing or decreasing we can calculate the probabilities for $h$ starting with $h'.$

For example, we can take $h'$ corresponding to the opposite complex structure. Then all values of "energy" change the sign. The ground state should be defined as the state with maximal "energy". However, physics does not change.

There exists a more invariant definition of ground state. First of all we define an equilibrium state as a robust stationary state (=robust zero mode of the "Hamiltonian") with maximal value of the entropy  for  fixed value of "energy". Then we can define the ground state as the equilibrium state with minimal entropy.

If two "energy" functionals are proportional they have the same set of equilibrium states, hence the same ground state.

The absence of natural energy functional sounds disturbing.  However,  in classical mechanics one can write down the equations of motion if we do not know the energy functional, but know the hypersurfaces of constant energy (this was noticed, for example, in \cite {ARNOLD} ). This means that we can replace the energy functional $h$ by $h'=f(h)$ where $f$ is increasing or decreasing,
precisely as in our picture.

More generally, if for an observable $(A,a)$ the numbers $a(z_k)$ are distinct the probabilities for the observable $(A,a')$ can be expressed in terms of probabilities for $(A,a)$ ( here $z_k$ are robust zero modes of $A$).

It is clear from the above considerations that the functional $a$ does not have a direct physical meaning; it can  be characterized as an accounting device. However, the probabilities $p_k$ make sense as physical quantities; they can be described as probabilities to find the state $x$ in pure states $z_k$ after interaction with environment.
  
 Define a projection $P:{\cal L}\to \cal L$ by the formula
 $$Px=\lim_{T\to\infty} \frac 1 T \int_0^T dt  \sigma_A(t)x$$
 where $\sigma_A(t)$ is the group of automorphisms generated by $A.$  
 It is easy to  check, that $ P\psi_j=0$ if $\epsilon_j\neq 0, P\psi_j=\psi_j $ if $\epsilon_j=0.$
 
 If all zero modes are robust then $P=P'$, hence we can calculate the probabilities using $P.$
 
 It is obvious that $P{\cal C}_0\subset {\cal C}_0.$  It is easy to check that generically all zero modes are robust; it follows that  $P'{\cal C}_0\subset {\cal C}_0,$ hence all probabilities $p_i$ are non-negative.

 We say that observables $(A_1,a_1),...,(A_n,a_n)$ are commuting if  $a_i$ is $A_j$-invariant for all $j$ and there exists a basis of the complexification of $\cal L$ consisting of  common eigenvectors of $A_1, ..., A_n$. Generalizing the above considerations we can define a joint probability distribution of commuting observables.
 
 We  gave a heuristic proof of decoherence for diagonalizable operators with discrete spectrum. One can consider also generalized diagonalizable operators  , defined as operators having a basis consisting of
  generalized eigenfunctions. More formally one can say that such a operator should be equivalent  (conjugate) to an operator of multiplication by a function in a space of functions on some measure space ( for example, on ${\mathbb {R}}^n).$
  
  Assuming that in (\ref {DI}) $j$ is a continuous parameter, $\psi_j$ is a generalized eigenvector  and imposing some additional conditions we can  modify our arguments to prove decoherence in this case.
 \section{Elimination of redundant states  }

Let us start with physical theory  based on the space of states (considered as a bounded
convex subset ${\cal C}_0$ of topological vector space $\cal L$) and a subgroup $\cal V$ of the group of automorphisms of  ${\cal C}_0$ (of bijective linear maps of $\cal L$ mapping ${\cal C}_0$ onto itself). An observable is defined as a pair $(A,a)$ where $A$ is an element of $Lie (\cal V)$ and $a$ is a linear functional on $\cal L$ obeying $a(Az)=0.$ We assume that $\cal V$ acts in natural way on observables ($A$ transforms according adjoint representation and $a$ transforms as a function on $\cal L$). We fix the set of observables.



We say that  there exist redundant states in the theory if on can find such states $x,y\in {\cal C}_0$ that for every observable $(A,a)$ we have $a(x)=a(y)$  (there are no observables that allow us to distinguish these states). In this case it is useful to work with theory without redundant states. To construct such a theory we in introduce an equivalence relation in $\cal L$ saying that $x\sim y$ if $a(x)=a(y)$ for every observable $(A,a).$ In the new theory
the set of states ${\cal C}'_0$ is defined as a set of equivalence classes in ${\cal C}_0.$
The group $\cal V$  acts on $\cal L'$ (on the space of equivalence classes in $\cal L$); its elements can be regarded as automorphisms of ${\cal C}'_0$. The observables descend to $\cal L'.$

Let us consider some examples.

Let us start with a symplectic manifold $M$  equipped with transitive action of  automorphism group $G$
(homogeneous symplectic manifold). (See, for example, \cite {SH} for basic facts of the theory of such manifolds.)  We assume that this action induces a homomorphism of the Lie algebra $\textgoth {g}$ of $G$ into Lie algebra of Hamiltonian  vector fields on $M$; the Hamiltonian function
 of the vector field corresponding to the element $X\in \textgoth {g}$ will be denoted by $H_X$. The Hamiltonian functions are specified up to  additive constants; we assume that these constants can be chosen in such a way that the map $X\to H_X$ is a homomorphism (it should transform the commutator of Lie algebra elements into Poisson bracket of  functions). Then one says that
 $M$ is a strictly symplectic homogeneous manifold. Orbits of the coadjoint action of the group $G$ 
 on the vector space $\textgoth {g}^*$ dual to the vector space $\textgoth {g}$ belong to this class.  (The space $\textgoth {g}^*$ is equipped with natural Poisson structure. Orbits are symplectic leaves of this structure. The symplectic structure on orbits is called Kirillov symplectic structure.)
 
 One defines the moment map $\mu$ of a  strictly symplectic homogeneous manifold $M$ into $\textgoth {g}^*$ as a map $x\to \mu_x$ where $\mu_x(X)=H_X(x)$ for $X\in \textgoth {g}.$  This map is $G$-equivariant with respect to coadjoint action of $G$ on $\textgoth {g}.$ The moment map is a local
symplectic  isomorphism of $M$  with one of orbits of coadjoint action. 
For every state of classical system  (for every probability distribution $\rho $ on $M$) we define a point $\nu(\rho)\in \textgoth {g}^*$  as an integral  of $\mu_x$ with respect to the measure $\rho$:
$$\nu(\rho)=\int_M \mu_xd\rho.$$
The point $\nu(\rho)$ belongs to the convex envelope $N$ of the orbit $\mu(M).$

The group $G$ acts naturally on the space of classical states. It follows from $G$-equivariance of the moment map that the map $\nu$ is a $G$-equivariant map of this space
into $\textgoth {g}^*$  equipped with coadjoint action of $G.$

We say that two classical states (two probability distributions $\rho$ and $\rho'$ ) are equivalent if 
\begin {equation}
\label {E}
\int H_X(x)d\rho= \int  H_X(x)d\rho'
\end {equation}
for every
 $X\in \textgoth {g}.$
In other words we say that two states are equivalent if  calculations with these states give the same results  for every Hamiltonian $H_X.$

We will derive the following statement:

{\it  Two states $\rho$ and $\rho'$  are equivalent iff $\nu(\rho)=\nu(\rho').$}

To give the proof we notice that for every $X\in \textgoth {g}$
$$\nu (\rho)(X)=\int_M  \mu_x(X)d\rho=\int_M  H_X(x)
d\rho$$
and similarly
$$\nu (\rho')(X)=\int_M  \mu_x(X)d\rho'=\int_M  H_X(x)
d\rho'.$$

The space of states of the classical theory with Hamiltonians restricted to the set $\{H_X\}$ where $X\in \textgoth {g}$ should be considered as the space of states on $M$  where equivalent states are identified (we eliminate redundant states).  The map $\nu$ is a bijective map of this space onto the set of states  $N$ obtained as a convex envelope of the orbit $\mu(M)$  ("quantum states").  The $G$-equivariance of the map $\nu$
means that the evolution of classical states agrees with the evolution of quantum states.

This statement can be used to construct physical  theories with any prescribed symmetry group.

Notice that our considerations can be applied to infinite-dimensional homogeneous symplectic manifolds.

Analogs of results of this section can be proved for homogeneous symplectic supermanifolds. The proofs are the same.

Notice that the requirement  that the group $G$ acts transitively on the manifold $M$ is not necessary in almost all of our considerations. It is used only to say that the image
of the moment map $\mu$ is an orbit of the coadjoint action.  For any strictly symplectic $G$-manifold  the image of the moment map $\mu$ is a union of orbits and the image of the map $\nu$ is a $G$-invariant convex subset  of $\textgoth {g}^*$  (the convex envelope of the image of $\mu$).

We obtain that eliminating redundant states in the  classical theory with Hamiltonians restricted to the set $\{H_X\}$  we get physical  theory with  the set of states $\nu(M)$ and the group $\cal V$ identified with the group $G$  acting on $\nu (M)$.\footnote {Notice that we assumed that the set of states of physical theory is bounded. This condition is not  always  satisfied for $\nu(M).$}  Observables of  the theory can be regarded as pairs $ (X,  H_X).$  One can prove that in this  theory  all zero modes of generic observable  are robust. Adding other Hamiltonians  to the Hamiltonians $H_X$ we obtain a classical theory that can be regarded as a deformation of  the theory we consider.

The complex projective space is   a strictly symplectic homogeneous $U$-manifold where $U$ denotes the unitary group.  Our constructions show that classical theory on  infinite-dimensional complex projective space with restricted set of Hamiltonians is equivalent to textbook quantum mechanics.  This statement is closely related to
 the constructions suggested by S. Weinberg \cite {WE} and the deformation of physical theory we mentioned is Weinberg's non-linear quantum mechanics. (I am indebted to A. Kapustin for this remark.) 
 
 Let us illustrate the above constructions in the case when $G$ is the group $U$ of unitary transformations of Hilbert space $\cal H.$ In this case we can identify  the elements of Lie algebra $\textgoth {g}$ with  self-adjoint  operators and the  dual space $\textgoth {g}^*$ with linear space of trace class self-adjoint operators.    ( In  the notations accepted in present paper  the elements of Lie algebra are  skew-adjoint operators; to identify them with self-adjoint operators  we multiply by $i.$)
  To simplify notations we assume that the Hilbert space is finite-dimensional , however our considerations can be applied also in infinite-dimensional case.  If 
 $\dim {\cal H}=n$ an orbit  is labeled by distinct real numbers $\lambda_1,..., \lambda_r$ (eigenvalues) and non-negative integers $k_1,...,k_r$
 obeying $k_1+...k_r=n$ (multiplicities of eigenvalues).  The stationary group of $U(n)$ -action on the orbit is isomorphic to the direct product of  groups $U(k_i)$, therefore the  orbit is homeomorphic to $U(n)/U(k_1)\times...\times U(k_r)$ ( to a flag manifold). If $r=2$ the orbit is homeomorphic to Grassmannian. If $r=2, k_1=n-1, k_2=1$ we obtain complex projective space.
 (The  Grassmannian  $G_k(\cal H)$  is defined as a space of all $k$-dimensional subspaces of $\cal H$; it  can be regarded as symplectic $U$-manifold. An orthonormal basis of $k$-dimensional subspace is defined up to a transformation from the unitary group $U(k)$; this means that  points of $G_k(\cal H)$  are described
 by  orthonormal systems of vectors $\phi_1,...,\phi_k$ with identification
 $\phi'_i\sim u_i^l\phi_l$ where $u_i^l$ is a unitary matrix.)
 
 Let us fix an orthonormal basis in $\cal H.$ This allows us to consider elements of $\textgoth {g}$ and $\textgoth {g}^*$ as Hermitian matrices.
 If a Hermitian matrix $X^a_b$ specifies an element of $\textgoth {g}$  the corresponding Hamiltonian function has the form $H_X (K)= X^a_bK^b_a$
 where $K$ is an element of $\textgoth {g}^*$ considered as a Hermitian matrix. In appropriate orthonormal basis in $\cal H$ the matrix $X^a_b$ is diagonal: $X^a_b=h^a\delta^a_b.$  In this basis all diagonal matrices  considered as elements of $\textgoth {g}^*$  are zero modes  of the action of $X^a_b$ (recall that  $\textgoth {g}$ acts $\textgoth {g}^*$ by means of coadjoint representation). If all diagonal entries $h^a$ are distinct all zero modes of $X^a_b$ are diagonal. In this case we can say that  all zero modes of $X^a_b$ acting on a convex envelope of an orbit are robust. Diagonal  matrices belonging to the orbit are pure robust zero modes.
 
 For arbitrary compact Lie group the coadjoint representation can be identified with adjoint representation. Without loss of generality we can
 assume that $X\in \textgoth {g}$ belongs to Cartan subalgebra $\textgoth{h}$. Elements of $\textgoth {h}$ belonging to an orbit are pure robust zero modes of corresponding physical theory.  If $X$ is a  regular element of Cartan subalgebra all zero modes of $X$ are robust.
 
Let us show that in geometric approach one can obtain any theory from classical theory  eliminating redundant states. ( We understand here classical theory as any theory where every state has a unique representation  as a mixture of pure states; in other words the set of states is a  Choquet  simplex.) Let us start with a theory with the set of states ${\cal C}_0$ and the set of observables $(A,a).$ We denote by 
$N$ the set of all pure states  and by  $\tilde {\cal C}_0$ the set of all probability distributions on $N.$  Elements of $\tilde {\cal C}_0$ can be considered as states of classical  theory
(it is clear that extreme points of this set can be identified with $N$.) The  observables of  classical theory by definition come from observables of the original theory.  Eliminating redundant states
in classical theory we come back to the original theory.

Using the remark that  classical theory is a particular case of physical theory where  every state can be represented uniquely as a mixture of pure  states it is easy to present classical theory as a limit of quantum theories with Planck constant $\hbar$ tending to zero. One of possible ways is based on the formalism of $L$-functionals suggested in \cite {SCH} ( see also \cite {S}, \cite {MO}). In this formalism we start with Weyl algebra ${\cal A}_{\hbar}$ defined as a unital associative algebra  generated by elements $a_k, a_k^+$ obeying canonical 
commutation relations (CCR) $$[a_k, a_l^+]=\hbar \delta _{k,l}, [a_k, a_l]= [a_k^+, a_l^+]=0.$$  
We consider Weyl algebra as an algebra with involution $^+.$ To every density matrix $K$  in representation space of Weyl  algebra (= space of representation of CCR)  we can assign a functional $L_K(\alpha^*,\alpha)$ defined by the formula
\begin{equation}
\label{l}
L_K(\alpha^*,\alpha)=Tr  e^{-\alpha a^+}e^{\alpha^*a}K
\end{equation}
Here $\alpha a^+$ stands for $\sum \alpha_k a^+_k$ and $\alpha^*a$ for $\sum \alpha^*_ka_k,$ where $k$ runs over  some set .
The functionals $L_K$ ($L$-functionals) can be considered as positive functionals (states) on the Weyl algebra ${\cal A}_{\hbar}.$ In the limit $\hbar\to 0$ they give positive functionals on commutative algebra (classical states). Equations  of motion for $L$-functionals have a limit as $\hbar \to 0$; in the limit we obtain classical equations of motion (see \cite {S} or \cite {MO} for more detail). This remark is especially useful in consideration of  quantum particles corresponding to  to  (generalized) solitons: their $L$-functionals have classical limit. The same is true  for the scattering matrix  of these particles in the  formalism of $L$-functionals (inclusive scattering matrix).

\section {Jordan algebras }
 Jordan algebra can be defined as a unital commutative algebra where the operators $R_x$ and $R_{x\circ x}$ commute. ( Here $R_u$ is an operator of multiplication by $u$, i.e. $R_u(v)=u\circ v.$)  A subalgebra of Jordan algebra generated by one element is associative ( the Jordan algebra is power- associative). This means that we can talk about powers $x^n$ of  an element $x.$ For every unital associative algebra we can  define a structure of Jordan algebra  introducing  the operation
$x\circ y=\frac 1 2 (xy+yx).$  One says that  a subalgebra of a Jordan algebra obtained this way is a  special Jordan algebra; algebras that are not special are called exceptional.

If a unital associative algebra is equipped with an involution the set of self-adjoint elements can be regarded as a Jordan algebra with respect to the operation $x\circ y.$

One can consider Jordan algebras over any field; for definiteness we consider Jordan algebras over $\mathbb {R}.$

 The formulation of physical theory in terms of the set  of states is closely related to the formulation in terms of Jordan algebras. For every Jordan algebra $\cal B$  we define a cone of positive elements  ${\cal B}_+$  as a convex  envelope of the set of  elements of the form $x^2$ where
 $x\in \cal B$. \footnote {  Notice that in our definition of cone  a vector space is also a cone} We can consider also the dual cone consisting of linear functionals on $\cal B$ that are positive on  ${\cal B}_+$. We can use one of these cones in geometric approach to physical theory.
 
 If $\cal B$ is a linear topological space and algebraic operations are continuous we say that
 $\cal B$ is a topological  Jordan algebra. For such algebras we consider only continuous functionals and maps.


We will consider $JB$-algebras $\cal  B$  defined as Jordan algebras that can be equipped with Banach norm obeying
$$ ||x\circ y||\leq ||x||\cdot ||y||, ||x^2||=||x||^2, ||x^2||\leq ||x^2+y^2||.$$ ( The first condition means  that $JB$-algebra is a Banach algebra. An associative Banach algebra obeying $||x^2||=||x||^2$ is called $C^*$-algebra. The set of self-adjoint elements of $C^*$-algebra is a $JB$-algebra with respect to the operation $x\circ y=\frac 1 2 (xy+yx).$  $JB$-algebras of this kind and their subalgebras are called $JC$-algebras.)

 Finite-dimensional $JB$-algebras coincide with Euclidean Jordan algebras classified by Jordan, von Neumann, Wigner.  They proved that almost all simple algebras of this type can be realized as
 algebras of Hermitian $n\times n$ matrices with real, complex, quaternionic or octonionic
 entries. (In octonionic case we should take $n=3$; we obtain 27-dimensional algebra called Albert algebra. The Albert algebra is exceptional.) There exists one more series of simple Euclidean algebras consisting of algebras with generators $1, e_1,...,e_n$ obeying relations $e_i\circ e_j=0$ for $i\neq j$, $e_i\circ e_i=1.$

We  defined  a positive cone $\mathcal {B}_+$ in any Jordan algebra $\cal B$ as a  convex  envelope of the set of  all squares.  In the case of $JB$-algebra one can say that the positive cone consists of squares. Equivalently  $a\in \mathcal {B}_+$ if the spectrum of $R_a$  consists of non-negative real numbers.

 If a $JB$-algebra comes from $C^*$-algebra this definition coincides with the definition of the positive cone in $C^*$-algebra (recall, that the positive cone in an associative algebra with involution is spanned by the elements of the form $A^*A$).


The group $\cal V$ can be defined as the group of automorphisms of the algebra $\cal B$ acting on the cone.
Its Lie algebra consists of derivations. One can consider also a larger group $\cal V$ consisting of all invertible structural transformations (structure group).  It is generated  by automorphisms and operators $Q_a=2R_a^2 -R_{a\circ a}$  where $a$ is invertible.

If the Jordan algebra $\cal B$ consists of self-adjoint elements of an algebra with involution $\cal A$  then every skew-adjoint element $T$ of $\cal A$  specifies a derivation $\alpha_T$  (as a commutator with $T$). An even polynomial  $p(T)$ is a self-adjoint element of $\cal A$ commuting with  $T$, hence it is a zero mode of the derivation $\alpha_T$. This  is a robust zero mode of the derivation:  a derivation $\alpha _{T'}$ where $T'$ is close to $T$ has a zero mode  $p(T')$ that is close to $p(T).$

There exists unique exceptional (not special) simple Jordan algebra. Any non-trivial  derivation of it has three robust zero modes. To prove this fact we realize this algebra as the algebra of $3\times 3$  Hermitian octonionic matrices.
Elements of the  group $SO(8)$ can be regarded as automorphisms of this algebra, elements of $so(8)=Lie SO(8)$ specify infinitesimal automorphisms having diagonal matrices as zero modes. These zero modes are robust. To prove this we notice that generic elements of $so(8)$ have only these three zero modes. From the other side all infinitesimal automorphisms can be transformed into elements of $so(8)$ by means of inner automorphisms of the automorphism group. This means that the every infinitesimal automorphism has at least three zero modes and generic infinitesimal automorphism has precisely
three zero modes. It is easy to conclude from this fact that these zero modes are robust.

Transformations  $$Q _a(x)=\{a,x,a\}=(2R_a^2 -R_{a\circ a})x$$ where
$$\{a,x,b\}=(a\circ x)\circ b+(x\circ b)\circ a-(a\circ b)\circ x$$
transform the cone into itself (belong to ${\rm End}\mathcal {B}_+.$)  If $a$ is invertible, then $Q_a$ is an automorphism of the cone. 

Noticing that $Q_a(1)=a^2$ we obtain that  the cone of $JB$-algebra is homogeneous (the automorphisms of the cone act transitively on the interior of the cone).
 If the algebra is finite-dimensional then the cone is self-dual and all self-dual homogeneous cones can be obtained this way. Therefore
 Jordan-von Neumann-Wigner theorem gives a  classification of finite-dimensional self-dual homogeneous cones (round cones, cones of positive self-adjoint operators in real, complex and quaternionic  vector spaces and the exceptional 27-dimensional cone).

All finite-dimensional homogeneous cones were described by E. Vinberg \cite {VIN}

It was conjectured that superstring is related to 
exceptional Jordan algebra (Foot-Joshi \cite {FJ}).
The group $SO(1,9)$ acts as a subgroup of the automorphisms of the cone  of this algebra.

We considered Jordan algebras over $\mathbb {R}.$ Complexifying these algebras we obtain 
Jordan algebras over $\mathbb {C}$ equipped with involution  (complex conjugation).
In particular, complexifying $JB$-algebras we obtain $JB^*$-algebras (this statement can be regarded as a definition of  $JB^*$-algebra, but there exists also an independent definition of this class of algebras).
\vskip .1in
\section {Geometric approach to quantum field theory. Particles and quasiparticles.}

In quantum field theory it is more convenient to work with the cone of non-normalized states ${\cal C}\subset {\cal L} $  where $\cal L$ is a Banach space or, more generally, topological vector space. Then the set of states ${\cal C}_0$ should be defined as the set of equivalence classes of  points of the cone with respect  to the equivalence relation $x\sim \lambda x.$  We define endomorphisms of the cone as linear operators on $\cal L$ transforming the cone into itself and commuting with multiplication by a  number. Automorphisms of the cone  are defined as bijective endomorphisms.

{\it The basic objects in our setting are the cone $\cal C$, a subgroup $\cal V$ of the group of automorphisms of the cone and a subsemiring $\cal W$  of the semiring of endomorphisms of the cone. }(Recall that the set of endomorphisms of the cone is closed with respect to addition and composition of operators as well as with respect to a multiplication by a positive number.  The set $\cal W$ also should be closed with respect to these operations.)  {\it We assume that the group $\cal V$ acts on $\cal W$ by conjugations ( i.e. for $v\in {\cal V}, w\in {\cal W}$ the operator $vwv^{-1}$ belongs to $\cal W$)}.\footnote {
Instead of taking the cone $\cal C$ as a starting point we could start with the semiring $\cal W$ and define the cone and the group $\cal V$ in terms of this semiring.}

To relate this setting to the picture of Section 2 we should assume that there exists a $\cal V$-invariant  linear functional $\alpha$ (normalizing functional)  such that  $\alpha(x)>0$ for every  non-zero element $x\in \cal C$. Then we can define  ${\cal C}_0$ as the subset of the  cone consisting of points obeying $\alpha (x)=1.$  In the considerations below we  do not need the  normalizing functional.

Notice that  one can  take  as  basic objects in geometric approach the cone $\cal C$ and  a subgroup $\cal V$ of the group of automorphisms of the cone ( without using the semiring $\cal W$). In this case one should use the second definition of excitation, that is more transparent, but less explicit. (The definitions of excitations are discussed below.)

Starting with an associative algebra $\cal A$ with involution $^*$  we define $\cal L$ as the set of linear functionals and the cone $\cal C$ as the set of positive linear functionals (functionals $f$ obeying $f(A^*A)\geq 0$). The group $\cal V$ is defined  as the group of automorphisms of  $\cal A$. The semiring $\cal W$
is generated by endomorphisms  $\Psi_ B$of $\cal C$ sending the functional $f(A)$ into the functional $f(B^*AB).$ These endomorphisms can be written in the form $\Psi_B=\tilde B B$ where
$(\tilde Bf)(A)=f(B^*A), (Bf)(A)=f(AB).$

Starting with a Jordan algebra $\cal B$ we can take as as $\cal L$ either $\cal B$ or the space of linear functionals on $\cal B.$   
 The cone $\cal C$   can be defined  either as the cone ${\cal B}_+$  of positive elements of $\cal B$ ( a convex envelope of all squares) or as a dual cone.  The  group $\cal V$ can be defined as the structure group or as the group of automorphisms of $\cal B.$ (The structure  semigroup  $Str(\cal B)$ is generated by automorphisms of $\cal B$ and  operators $Q_a= 2R_a^2 -R_{a\circ a}$  where $R_a$ stands for Jordan multiplication by the element $a\in \cal B.$  Requiring  that $Q_a$ are invertible we get a definition of structure group.) The semiring $\cal W$ can be defined as the  semiring generated by operators $Q_a.$

Starting with a contact $G$- manifold $M$ we can define $\cal C$ as as the moment cone. Recall that a contact  structure is specified by a non-degenerate one-form $\alpha$ (contact one-form) defined up to multiplication by a positive function.  We say that  $M$ is a contact $G$-manifold if the group $G$  acts on $M$ by transformations preserving contact structure  (i.e. they transform  a  contact one-form into
a form specifying the same contact structure). For a contact form $\alpha$ we define the $\alpha$-moment map $\mu_{\alpha}:M\to \textgoth {g}^*$ where $\textgoth {g}$ stands for the Lie algebra of $G$ by the formula
$$\langle\mu_{\alpha}(m), X\rangle=\langle \alpha, Xm\rangle$$
where $m\in M, X\in \textgoth {g}.$ The set $\mu_{\alpha}(M)$ depends on the choice of the contact form $\alpha$, but  the set  $\cal M$  of points
of the form $\rho x$ where $\rho\geq 0, x\in \mu_{\alpha}(M)$ depends only on contact structure. This follows immediately from the formula 
$$\mu_{f\alpha}(m)= f(m)\mu_{\alpha}(m).$$
We say that $\cal M$ is the moment "cone".( The quotation marks are necessary, because  in our definition a cone is a convex set.) \footnote {One can construct  the moment "cone" using symplectization  of a contact manifold and symplectic moment map (see, for example, \cite {AG} for the notion of symplectization). Notice that we assumed the existence of a  global contact form in the definition of contact manifold; in mathematical terminology this means that we consider co-orientable contact manifolds.}

We can define $\cal C$ as a convex envelope of $\cal M$ (and  call it the moment cone). The group $G$ acts on $\cal C$, hence we can define $\cal V$ as $G.$


 In geometric approach  (quasi)particles and the scattering of (quasi)particles can be defined if 
 an  abelian Lie group interpreted as a group of space-time translations acts on the cone of states ${\cal C}.$  The translations should belong to the group $\cal V.$
We denote spatial translations by $T_{\bf x}$  where ${\bf x} \in \mathbb {R}^d$ and time translations by $T_{\tau}.$  We assume also that translations act also on  $\cal W$ and this action is compatible with the action on the states. We  use the notation $A(\tau, {\bf x})$ for the translated operator $A.$

In Lorentz-invariant theory the action of translations can be extended to the action of Poincar\'e group $\cal P.$

In geometric approach an excitation of translation-invariant stationary state $\omega$ can be defined as a state of the form $W\omega$ where $W\in \cal W.$  ( We assumed that $\cal W$ is a semiring, therefore the set of excitations ${\cal W}\omega$ is a cone.) Alternatively one can say that an  excitation is  a  state $\sigma$ obeying $T_{{\bx}}\sigma\to C\omega$ as ${\bx}\to \infty$ (here $T_{{\bx}}$ stands for spatial translation, C is a constant factor).   The second definition is the most transparent one. One can say not very precisely  that the excitation essentially differs from $\omega$ only in a bounded spatial domain.

 Let us establish the  relation between two definitions in algebraic quantum field theory. Recall that in this case $\cal W$ is the smallest semiring containing elements of the form $\tilde B B $ where $B\in \cal A.$  Hence to prove that an excitation in the first sense is an excitation in second sense one should check that the state $\sigma(A)=\omega (B^*AB)$ obeys the conditions of the second definition. We assume that $\omega$ obeys the cluster property.  
This means, in particular, that
$$\omega(B^*A(\tau,{\bx})B)-\omega(B^*B)\omega(A(\tau,{\bx}))\to 0$$
as $\bx\to\infty$. Using translation invariance of $\omega$ we obtain that  in this limit
$T_{{\bx}}\sigma\to C\omega$ with $C=\omega (B^*B).$

Notice that  the above proof can be used to show that every state $\sigma (A)$  that can be represented by the formula $\sigma (A)=\omega (B'AB)$
 where $B,B'\in \cal A$ obeys the conditions of the second definition.  If $B'=B^*$ this formula always specifies a state; in general this is wrong.  However, in the case when $\omega$ lies in the interior of the cone and $B,B'$ are close to  the unit element of the algebra $\cal A$ the functional $\sigma$ also lies in the cone (=specifies a state).
 We see that the second  definition is broader than the first one.
 
 Notice that the second definition of excitation does not depend on the choice of the semiring $\cal W.$ In what follows we can use either  first or 
second definition. 

 Quasi-particles can be defined as elementary excitations of  translation-invariant stationary state $\omega.$

Particles are defined as elementary excitations of ground state.

To make these definitions precise we should  explain the notion of elementary excitation. We start with the explanation in the algebraic approach to quantum theory. In this approach  the action of translations on states is induced by the action of translations on the algebra $\cal A$. The time and spatial translations are defined as involution-preserving  automorphisms  $\alpha (\tau, {\bx}) $; we use the notation $A(\tau, {\bx})= \alpha (\tau, {\bx})A$ for $A\in \cal A$. The  GNS  ( Gelfand-Naimark-Segal)  construction gives a representation  $A\to \hat A$ of the algebra $\cal A$ in the pre Hilbert space $\cal H$ and a cyclic vector $\theta$ corresponding to the state $\omega$ (i.e. obeying $\omega (A)=\langle \hat A\theta, \theta\rangle$).  The translations descend to the  space $\cal H$ as unitary  (or orthogonal) operators $T_{\tau}, T_{{\bx}}$ (this follows from our assumption  that $\omega$ is a stationary translation-invariant state). Namely, we define define $T_{\tau}\hat  A\theta$ as $\widehat {A(\tau,0)}\theta$, $T_{\bx}\hat A\theta$ as $\widehat {A(0,{\bx})}\theta.$  Notice that $\widehat {A(\tau, {\bx})}=T_{\tau}T_{\bx}\hat A T_{-{\bx}}T_{-\tau}. $ The operators of energy and momentum $\hat H,\hat {\bP}$ are defined as infinitesimal translations (if we are working in real Hilbert space, they act in its complexification.).  We say that the states corresponding to  the elements of $\cal H$  are excitations of $\omega.$ This definition agrees with the definition in geometric approach:  if $\Theta=B\theta$ and $\sigma$ denotes the state corresponding to $\Theta$ then
$$\sigma (A)=\langle A\Theta, \Theta\rangle=\langle B^*AB\theta,\theta\rangle=\omega (B^*AB)=(W\omega)(A)$$
where $W=\tilde B B\in\cal W$.)

In Lorentz-invariant theory the Poincar\'e group $\cal P$ acts as a group of automorphisms of the algebra $\cal A$. This action induces an action of $\cal P$ on  states   and  a unitary (or orthogonal ) representation of this  group on the space $\cal H.$  An elementary excitation can be defined as an irreducible subrepresentation of  this representation.

Notice that we consider $\cal H$ as a pre Hilbert space; by definition a unitary representation in pre Hilbert space is irreducible if  it  induces an irreducible representation in the completion. An irreducible unitary representation in Hilbert space is isomorphic to the representation in  the space $L^2$ of square integrable functions , a representation in pre Hilbert space is isomorphic to the representation in a dense subspace of $L^2.$

An irreducible unitary representation of Poincar\'e group  with positive energy is  isomorphic to a representation of this group in the space of (multicomponent) functions  depending on the momentum $\bf k$; the momentum operator $\hat \bP$ can be represented as a multiplication by $\bf k$ and the translation  $T_{\bx}$ is an operator of multiplication by $ e^{i\bx\bk}.$  This fact prompts the definition of  elementary excitation in general case: we assume that the representation of the group of spatial translations is the same as in Lorentz-invariant situation.

Let  us consider   an algebra over complex numbers  $\cal A$, a stationary translation invariant state  $\omega$, a complex pre Hilbert space $\cal H$  and a vector  $\theta\in \cal H$ obtained from $\omega$ by means of GNS construction. Then the
elementary excitation can be defined as  a generalized multicomponent  function $\Phi ({\bk})=(\Phi_1({\bk}), \cdots \Phi_m(\bk))$ such that  $\hat {\bP} \Phi ({\bk})={\bk} \Phi ({\bk})$, $\hat H \Phi ({\bk}) =E (
{\bk})\Phi ({\bk})$ where $E(\bk)$ is a matrix function taking values in Hermitian matrices.

We assume that $\Phi ({\bk})$ takes values in $\cal H$ and is delta- normalized: $\langle \Phi ({\bk}),\Phi ({\bk}')\rangle=\delta ({\bk}-{\bk}').$  In other words,  we have a linear operator  $\phi \to \Phi (\phi)$ that assigns to  every  $\phi\in \textgoth {h}$ a vector $\Phi (\phi)=\int \phi ({\bk})\Phi ({\bk}) d{\bk}$ in $\cal H.$ This operator should be an isometry obeying
$\hat H \Phi (\phi)=\Phi (\hat E \phi), \hat {\bP}\Phi(\phi)=\Phi(\hat {{\bk}}\phi)$ where $\hat E$ and $\hat {{\bk}}$ stand for multiplication operators by $E ({\bk})$ and ${\bk}.$  Here we take as the space $\textgoth {h}$ the space of complex square-integrable functions on  $\mathbb {R}^d\times \cal I$ ( here $\cal I$ is a finite set  consisting of $m$ elements) or any dense linear subspace of this space that is invariant with respect to the operators $\hat E$ and $\hat {{\bk}}.$  (In other words, these functions depend on the momentum variable ${\bk}\in \mathbb{R}^d$ and discrete parameter $j\in \cal I$. For definiteness we assume that these functions belong to the space $\cal S$ of smooth fast decreasing functions  ( all of their derivatives should tend to zero faster than any power).)  We can work also in coordinate representation assuming that the momentum operator $\hat \bP=\frac 1 i \nabla$ is the infinitesimal spatial translation (spatial translations are represented as shifts with respect to the coordinate variable $\bx$). 

We say that $\textgoth {h}$ is an "elementary space" over $\mathbb {C}.$


If $\cal A$ is an  algebra  over real numbers with action of spatial and time translations we can define the elementary excitations of translation-invariant stationary state $\omega$ in the following way.

Let us fix the space $\textgoth {h}$  as a subspace of the space of real square-integrable functions on $\mathbb {R}^d\times \cal I$ where $\cal I$ is a finite set. For definiteness {\it  we take $\textgoth {h}$ as the space $\cal S$ of smooth fast decreasing functions of $\bx$}. Then $\textgoth {h}$ is invariant with respect to spatial translations $T_{\ba}:\phi ({\bx},j)\to \phi ({\bx}-\ba,j)$; we assume that it is invariant with respect to time translations  (one-parameter group of orthogonal  operators commuting with spatial translations). The time translations can be written in the form $T_{\tau} =e^{-\tau\hat E}$ where $\hat E$ is a skew-adjoint operator with translation-invariant kernel. (In other words the kernel of the operator $\hat E$ has the form $E_{ab}(x-y)$ where $E_{ab}(x)=- E_{ba}(-x), a,b\in \cal I$.) We say that $\textgoth {h}$ is an {\it "elementary space" over $\mathbb {R}$}. 
\begin {definition}
An elementary excitation  of a stationary translation-invariant state $\omega$  is  an isometric map $\Phi$ of $\textgoth {h}$ into the space $\cal H$ of the corresponding  GNS -representation such that the translations in $\textgoth {h}$ agree with translations in $\cal H$ (i.e. $T_{{\bx}}\Phi (\phi)= \Phi (T_{{\bx}}\phi),
T_{\tau}\Phi (\phi)=\Phi (T_{\tau}\phi$).
\end {definition}

We formulated this definition for the case when $\cal A$ is an algebra over real numbers, but it can be applied also in the case when $\cal A$ is an algebra over $\mathbb {C}.$

 Considering the elements of $\textgoth {h}$ as test functions we can say  that elementary excitations are generalized functions $\Phi ({\bx}, j)$ taking values in $\cal H$ (here $j$ is a discrete index: $ j\in \cal I$). (We define the generalized function by the formula $\Psi (\phi) =\sum_j\int d{\bx} \Phi ({\bx}, j)\phi ({\bx}, j).$)

One can work in momentum representation. Then the test functions depend on the momentum variable ${\bk}\in \mathbb {R}^d$ and discrete variable $j\in \cal I$; if the test functions in coordinate representation are real then the test functions in momentum space obey the condition   $\phi ^*({\bk},j)=\phi (-{\bk},j).$  The spatial translation 
$T_{\ba}$ can be understood as multiplication by $e^{i\ba {\bk}}$.  A time translation acts as  multiplication by a matrix function $e^{i\tau E({\bk})}$ where ${\bk}\in \mathbb {R}^d
$ and $E({\bk})$ is a Hermitian  $(m\times m)$ -matrix. (Here $m$ is the number of elements in $\cal I.$)    Diagonalizing  $E(\bk)$ we can calculate the matrix function  $e^{i\tau E({\bk})}$; it has the form
\begin {equation} \label {TT}
 e^{i\tau E({\bk})}= \sum_ja_j({\bk})e^{i\epsilon_j({\bk}) \tau}
 \end{equation}
 where $ \epsilon_j(\bk)$ are eigenvalues of $E(\bk)$ and $a_j(\bk)$ are matrix functions.
 Notice  that $E(-{\bk})= -E(\bk)$ if the test functions are real.
 
 As usual the generator of time translations is identified with the observable corresponding to energy. The corresponding functional $h$  (the energy functional) can be chosen in the form
 $h(K)=Tr \hat B\hat E K$ where $\hat B$ is any skew-adjoint translation-invariant operator in $\textgoth {h}$ commuting with $\hat E.$ Notice that in the case when $\textgoth {h}$ is an "elementary space" over $\mathbb C$ represented as an "elementary space" over $\mathbb R$ there exists a natural choice of $\hat B$ as an operator corresponding to the multiplication by $i.$

 The formula  (\ref {TT}) allows  us to analyze the asymptotic behavior of $T_{\tau}$ as $\tau\to \infty$  in coordinate representation.

   Let us  denote by $U_{\phi}$  an open subset of  $\mathbb {R}^d$ containing all points  having the form $ \nabla \epsilon_s ({{\bk}})$ where ${\bk}$ belongs to $\rm {supp}(\phi)=\cup_j \rm {supp}\phi _j)$ (to the union of supports of the functions $\phi (\bk,j)$).

\begin {lemma} \label {K}
 {Let us assume that   $\rm {supp}(\phi)$ is a compact subset of $\cal R.$ Then  for large $|\tau|$ we have
$$ |(T_{\tau}\phi)({\bx}, j)|<  C_n (1+|{\bx}|^{2}+\tau^2)^{-n}$$
where $ \frac {{\bx}}{\tau}\notin U_{\phi}$, the initial data $\phi=\phi ({\bx},j)$ is the Fourier transform of $\phi ({\bk},j)$, 
and $n$ is an arbitrary integer.}
\end {lemma}
The proof of this lemma can be given by means of the stationary phase method.

We can express Lemma \ref {K} saying that $\tau U_{\phi}$ is an essential support of $(T_{\tau}\phi)({\bx}, j)$  for large $|\tau|.$

Let us consider now physical theories in geometric approach. To define elementary excitations we need the action of spatial and time translations on the cone $\cal C.$

\begin {definition}
In geometric approach  we define an elementary excitation of translation-invariant stationary state $\omega$ as a  map of $\textgoth {h}$ into the set of excitations of $\omega$. This map should agree with the action of spatial and time translations.
\end{definition}
To relate this definition to the definition of elementary excitations in algebraic approach we  notice that starting with a map $\Phi:\textgoth {h}\to \cal H$ specifying an elementary excitation we can construct a  quadratic map $\sigma$ sending $\phi\in \textgoth {h}$ into a  state  $\sigma_{\phi}$ defined by the formula $\sigma_{\phi}(A)= \langle A\Phi(\phi), \Phi (\phi)\rangle.$ ( Here a state is a positive linear functional on $\cal A$ where $\cal A$ is an algebra with involution over $\mathbb {R}$.) 

Staring with real pre Hilbert space  $\textgoth {h}$ we can construct a cone ${\cal C} (\textgoth {h})$ as a convex envelope of points of the form $x\otimes x$ in the tensor square of  $\textgoth {h}$.  ( For every real Hilbert space $R$  the points of the form $x\otimes x$ in the tensor square of $R$ correspond to extreme points of the cone positive definite trace class 
operators in $R$.)  A linear map of the cone ${\cal C} (\textgoth {h})$ into the set  of excitations can be regarded as quadratic map of $\textgoth {h}$
into this set.

Similar constructions work for complex spaces, but instead of points of the form $x\otimes x$ in tensor square of $\textgoth {h}$ we should work with points of the form $x\otimes \bar x$ belonging to the tensor product of $\textgoth {h}$ and complex conjugate space $\overline {\textgoth {h}}.$
The linear envelope of these points is a cone denoted by ${\cal C} (\textgoth {h}).$

A linear map $l$ of tensor square defines a quadratic map $q$ by the formula $q(x)=l(x\otimes x).$  We say that a linear map $l$ of tensor product of complex vector space and complex conjugate space defines a  Hermitian map $q$ by the formula $q(x)=l(x\otimes \bar x).$  

 A linear map of the cone ${\cal C} (\textgoth {h})$ into the set  of excitations can be regarded as a Hermitian map of $\textgoth {h}$
into this set. If this map commutes with translations it specifies an elementary excitation.

Notice that in scattering theory \cite {GA3} we impose some additional conditions on  elementary excitations of $\omega$ in geometric approach. 

If we are starting with classical  field theory in Hamiltonian or Lagrangian approach then the classical vacuum can be regarded as stationary translation-invariant field configuration with minimal energy density.
We can consider  "excitations" of translation-invariant field  as fields having finite energy or as fields that coincide with translation-invariant field at spatial infinity. ( Talking about the energy of an excitation we assume that that energy of translation-invariant field is equal to zero.) All excitations of classical vacuum should have non-negative energy. 

Quantizing classical field theory we expect that the ground state (physical vacuum) is obtained from the  classical vacuum and that the quadratic part of the action functional in the neighborhood of classical vacuum governs 
the excitations of ground state  (quantum particles).  The quantum particles corresponding to the quadratic part of action functional are called elementary particles.  However, it is possible that there exist other (composite) particles. Especially interesting particles correspond to solitons ( to finite energy solutions to the classical equations of motion having the form $s({\bx} -{\bv}t)$). Usually we have a family of solitons labelled by momentum $\bp$ (in Lorentz-invariant theories this is always the case). Then the set of fields
$s_{\bp}({\bx}-{\bf a})$  is a symplectic submanifold   of phase space that is invariant with respect to spatial and time translations. The restriction of the Hamiltonian to this manifold has the form $H({\bp},{\bf a})= E(\bp).$
Quantizing this manifold we obtain a quantum particle that in zeroth order with respect to $\hbar$  corresponds to "elementary space" over $\mathbb {C}$ with the set $\cal I$ consisting of one element  ($m=1$) and with infinitesimal time translation  governed by the function $E(\bp).$ Generalized solitons also correspond to symplectic submanifolds of phase space that are invariant with respect to spatial and time translations; after quantization they lead to quantum particles described  by "elementary spaces" over $\mathbb {C}$  with $m>1.$  The classical limit of states of these quantum particles can be understood in the language of $L$-functionals ( see
Section 3).

Finally a remark about the elementary excitations in the formulation in terms of Jordan algebras. We assume that  time and spatial translations act as automorphisms of $JB$-algebra; then they act also on the positive cone and on the dual cone.

Let us fix a linear map $\rho: \textgoth {h}\to \cal B$ commuting with translations. Using the quadratic map $Q: \mathcal{B}\to  End (\cal{B}_+)$  we can define a quadratic map of $\cal B$  into the the space of excitations of translation-invariant stationary element $\omega\in \mathcal {B}_+.$ Then the composition of this map with $\rho$ gives an elementary excitation of $\omega$  (it sends $\phi\in \textgoth {h}$ into $Q_{\rho (\phi)}\omega$). This follows from the fact that $Q$ commutes with automorphisms (hence with translations).

There exists a similar construction of elementary excitations in the case when we are working with the dual cone .

The above considerations can be generalized to the case when translations act  by elements of the structure group.  Recall that the structure group $\rm {Str g} (\cal B)$ is generated by automorphisms and invertible quadratic maps $Q_a$.  The structure semigroup  $\rm {Str}(\cal B)$ is generated by all $Q_a$ and automorphisms. The elements of the structure semigroup are called structural transformations.   We define an involution $A\to A^t$  on the structure group and structure semigroup assuming that  it transforms $Q_a$ into itself and transforms an automorphism into an inverse automorphism.

The structure group acs by means of automorphisms on the cone $\cal B_+$ and on the dual cone $\cal B^*_+$.
The structure semigroup acts by endomorphisms of the cones. 

The map $Q_a$  agrees with structural transformations $B\in {\rm Str} (\cal B)$ in the following sense:

\begin {equation}
\label {Q}
Q_{Ba}= BQ_aB^t
\end {equation}

Let us consider as an example a  $JB$-algebra $\cal B$ of self-adjoint elements of $C^*$-algebra $\cal A.$
In this algebra every element $A\in \cal A$ specifies a structural transformation $x\to AxA^*.$  If $A$ is self-adjoint this transformation coincides with $Q_A$, if $A$ is orthogonal (or unitary in complex case) then this transformation is an automorphism.

Let us denote the translation group by $\cal  T.$ As usual we use the notation $T_{\tau}T_{\bx}A=A(\tau,\bx).$ We assume that the involution $A\to A^t$ transforms  translations into translations.  Let us denote by $\omega$ a translation invariant element of the cone or  of the dual cone. It follows from (\ref {Q}) that 
\begin {equation}\label {QQ}
Q_{a({\tau,\bx})}\omega =T_{\tau}T_{\bx}Q_a\omega.
\end{equation}

As we noticed starting with a linear map $\rho: \textgoth {h}\to \cal B$ commuting with translations  we can define a quadratic map of $\textgoth {h}$  into  the cone of excitations of translation-invariant stationary element $\omega\in \mathcal {B}_+$ or $\omega\in \mathcal {B}_+^*$ as a composition of the map $a\to Q_a\omega$ with $\rho$. To prove that this map  gives an elementary excitation of $\omega$ we should check that it agrees with the action of translations; this follows from (\ref {QQ}).

Our considerations can be repeated in  the case when $\textgoth {h}$ is a complex elementary space and $\cal B$ is a $JB^*$-algebra. In this case we define a Hermitian map
of $\textgoth {h}$ into the cone of excitations of translation-invariant stationary element $\omega$  in the positive cone of $\cal B$ as a composition of the map $a\to \{a,\omega,a^*\}$ 
with a map $\rho: \textgoth {h}\to \cal B$ commuting with  translations.  (We are using the fact
that  $ \{a,\omega,a^*\}=(Q_{\alpha}+Q_{\beta})\omega$ where $\alpha$ stands for the real part of $a$ and $\beta$ stands for the imaginary part of $a$.) 

Similar constructions work  for
 translation-invariant stationary element $\omega$ of the dual cone.

{\bf Acknowledgements} I am indebted to  M. Douglas, Ya. Eliashberg,  D. Fuchs, A. Givental, A. Kapustin,  A. Kirillov, A. Konechny,  A. Mikhailov,  A. Polyakov, A. Rosly,  Yu. Suhov and A. Vainshtein for valuable discussions and important comments.

\end {document}